\documentclass[doublecol]{epl2}

\usepackage{amsmath}

\title{Optimal transport of ultracold atoms in the non-adiabatic regime}

\author{A. Couvert\inst{1} \and  T. Kawalec\inst{1} \and  G. Reinaudi\inst{1} \and  D. Gu\'ery-Odelin\inst{1,2}}
\shortauthor{A. Couvert \etal}

\institute{
  \inst{1} Laboratoire Kastler Brossel, CNRS UMR 8852, Ecole Normale
Sup\'erieure, 24 rue Lhomond, 75005 Paris, France\\
  \inst{2} Laboratoire Collisions Agr\'egats R\'eactivit\'e, CNRS UMR 5589, IRSAMC,
  Universit\'e Paul Sabatier, 118 Route de Narbonne, 31062 Toulouse CEDEX 4, France
}


\abstract{We report the transport of ultracold atoms with optical
tweezers in the non-adiabatic regime, i.e. on a time scale on the
order of the oscillation period. We have found a set of discrete
transport durations for which the transport is not accompanied by
any excitation of the centre of mass of the cloud after the
transport. We show that the residual amplitude of oscillation of
the dipole mode is given by the Fourier transform of the velocity
profile imposed to the trap for the transport. This formalism
leads to a simple interpretation of our data and simple methods
for optimizing trapped particles displacement in the non-adiabatic
regime.}

\begin{document}

\maketitle

The controlled transport of ultracold atoms is crucial for the development of experiments in atomic physics. It makes possible
the delivery of cold atoms in a region free of the laser beams and coils of the magneto-optical trap (MOT), allowing a better optical and mechanical access. It also opens new perspectives for probing a surface or any material structure, and for loading atoms in optical lattices, or for positioning atoms in a high-Q optical cavity~\cite{SFC04,CSD07}. In addition it opens the way to a new generation of experimental setups where ultracold clouds of atoms would be delivered on demand on a variety of different
experimental platforms separated by macroscopic distances. This is standard for charged particles and energetic neutral particles,
while it has only been recently accomplished with ultracold atoms by moving slowly optical tweezers~\cite{GCL02}. Transport of
cold packets of atoms is also of importance as a step towards the continuous production of a Bose-Einstein condensate~\cite{CSL02,LRW06}.

Macroscopic transport of cold atoms has been demonstrated using several different configurations. One can move mechanically a pair of coils~\cite{LHW03,NSH05} or use a set of coils with time-varying currents~\cite{GBH01}. Such quadrupolar traps are non-harmonic. Alternatively one can use traps with a harmonic shape near their bottom, such as Ioffe-Pritchard traps~\cite{HRH01,HHS05,GKK05,LRW06}, optical tweezers as recently demonstrated on Bose condensed clouds~\cite{CSL02} or 1-D optical lattices~\cite{KAS01,STW06}. The harmonic potential is of particular interest since the centre of mass motion (also referred as the Kohn's mode) is not coupled to the other degrees of freedom, and this is true both in presence and absence of interactions between atoms and both for classical and quantum physics. However all these studies have been performed in the adiabatic regime where the duration of the transport is long with respect to the typical oscillation period of the trapped atoms. This is because a lot of energy is given to the trapped cloud when it is transported in the non-adiabatic regime, giving rise to heating and to a strong excitation of the dipole mode. This, in turn, can result in atom losses due to the finite depth of the trap. While microtraps can have high oscillation frequencies, the traps allowing to transport a large number of atoms are not very steep and thus an adiabatic transport is quite long, limiting the repetition rate of the experiments performed. To our knowledge, the issue of an optimal transport beyond this limit has only been addressed numerically for ions in Paul traps~\cite{SPS06}.

In this letter we report the transport of a cold atom cloud in the non-adiabatic regime with a high degree of control by means of optical tweezers with no residual excitation of the dipole mode of oscillation, moderate heating and no losses. We also provide a simple theoretical model which permits to work out a new picture of the transport. The residual amplitude of oscillation of the cloud can be expressed as the Fourier transform of the velocity profile imposed to the trap, yielding a simple interpretation of our data and providing simple methods for optimizing trapped particles displacement.

\begin{figure}
\onefigure{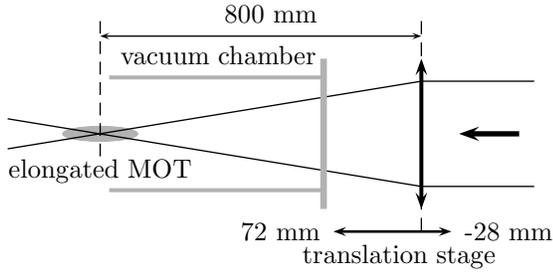} \caption{Sketch of the main part of the
experimental setup (not to scale) --- see text.} \label{setup}
\end{figure}

Our optical tweezers are generated by an Ytterbium fibre laser (IPG LASER, model YLR-300-LP) with a central wavelength of $1072$~nm. The wavelength of the laser is larger than the atomic resonance wavelengths of $780.24$~nm and 794.98 nm of the Rubidium~$87$ atoms, and thus, atoms are attracted to the region of maximum intensity~\cite{GWO00}. The beam is focused inside the vacuum chamber by a lens with a $802$~mm focal length mounted on a translation stage (Newport linear motor stage, model XMS100), allowing one to move the optical tweezers longitudinally on a $100$~mm range with an absolute repeatability on the order of a few hundreds of nm (see fig.~\ref{setup}). The resulting waist has been measured to be $44$~$\mu$m, corresponding to a Rayleigh length of $5.7$~mm.

The optical tweezers are loaded from an elongated MOT. The cigar shape of the MOT results from the two-dimensional magnetic gradients: $(0, 5, -5)$ G/cm. To maximize the loading of atoms into the dipole beam, the optical tweezers are superimposed on the MOT along its long axis. In addition, we favorize the selection of atoms in the hyperfine low level $5S_{1/2},F=1$ by removing the repump light in the overlapping region similarly to the dark MOT technique~\cite{KDJ93}.

The dipole trapping beam is turned on at a power of $80$~W during the 500 ms loading time of the MOT. Then, we increase the MOT detuning in 5 ms from $-3\Gamma$ to $-7.7\Gamma$, $\Gamma$ being the natural frequency width of the excited state. This procedure improves significantly the optical tweezers loading efficiency. Then, the magnetic field and repumping light are switched off to optically depump atoms to the $F=1$ ground sublevel. Finally all the remaining MOT beams are turned off. The number of atoms in the optical tweezers is as high as $3\times10^{7}$ corresponding to a peak atomic density of $5\times 10^{12}$~at/cm$^{3}$. These numbers are measured $50$~ms after switching off the MOT beams, so a first evaporation has already occurred on this time scale since the collision rate is larger than 500~s$^{-1}$.

In order to transport a cloud in the non-adiabatic regime without spilling atoms, one has to maximize the parameter $\eta = U_0/k_BT$ which is the ratio between the potential well depth $U_0$ and the average potential energy $k_BT$. We proceed in two steps. First, we cool down the sample by forced evaporation by lowering the beam power $P$. During this whole phase $\eta$ remains roughly constant. Second, we adiabatically re-compress the trap by increasing the beam power $P$. In this process, $U_0$ scales as $P$ and the temperature $T$ scales as $P^{1/2}$, and thus the dimensionless parameter $\eta$ increases as $P^{1/2}$. This way we can control the value of $\eta$ for a given power $P$ after compression.

Two different atom cloud preparation schemes were used. In the first one, referred to scheme 1, the initial trapping beam power is lowered in two linear ramps by a factor of $23$ within $600$~ms. The atomic cloud temperature before re-compression is $27\pm1.0$~$\mu$K. In scheme 2 the beam power is decreased in four linear ramps by a factor of $170$ within $3300$~ms, resulting in a $3.7\pm0.5$~$\mu$K temperature of the atomic packet. The trapping beam power after compression and before transporting the atoms reaches $37$~W (resp. $42$~W), and the temperature of the transported packets is $160\pm11$~$\mu$K (resp. $43\pm2$~$\mu$K) for scheme 1 (resp. 2). The $\eta$ parameter is thus equal to $13$ for scheme 1 and $50$ for scheme 2. The initial number of atoms before the transport is $2.1\times10^{6}$ (resp. $5.7\times10^{6}$) for scheme 1 (resp. 2).

The radial angular frequencies of the recompressed trap were inferred from a parametric heating experiment, and are on the order of $2$ kHz for both schemes. The longitudinal angular frequency was measured by examining the cloud dipole mode oscillations. We find $\omega_0=2\pi\times(8.1\pm0.3)$~Hz (resp. $\omega_0=2\pi\times(8.9\pm0.3)$~Hz) for the scheme 1 (resp. 2).

\begin{figure}
\onefigure{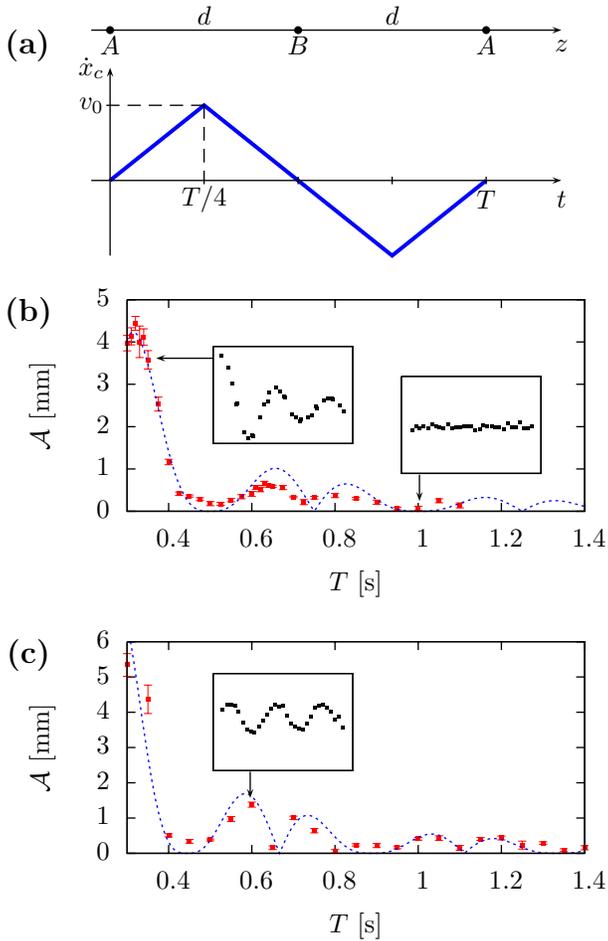} \caption{(Colour online) (a) Velocity profile
imposed to the trap to do the back and forth transport between $A$
to $B$ separated by $d=22.5$ mm. (b) (resp. (c)) The measured
amplitude $\mathcal{A}$ of the centre of mass dipole oscillation
for the conditions of scheme 1 (resp. 2), see text. The dashed
line is the theoretical prediction of eq.~\eqref{eq.ABA} with the
measured angular frequency $\omega_0$ of the trap.} \label{fig.2}
\end{figure}

The transport experiment has been carried out in a single vacuum
chamber. We consequently imposed a ``round trip'' displacement to
the optical tweezers, going from the MOT location $A$ to a point
$B$ placed $d=22.5$~mm from it along the beam direction, and back
to $A$ (see fig.~\ref{fig.2}a). The velocity of the trap as a
function of time is deliberately chosen as a succession of
constant acceleration for sake of simplicity. First, the trap is
accelerated at a constant rate $a$ during a time $T/4$,
decelerated at the opposite rate $-a$ during $T/2$, and finally
re-accelerated at $a$ to stop after a total transport time $T$.
This simple velocity profile will allow us to exhibit the main
features of non-adiabatic transport, and hence does not restrict
the generality of the conclusions that we will draw from our
experiments. The distance of transport $2d$ is simply related to
the acceleration $a$ and the transport duration $T$ by
$2d=aT^2/8$. In practice, the different transport durations that
have been used were obtained by varying the acceleration $a$ from
$0.2$ to $3.3\un{ms^{-2}}$, in order to investigate the
non-adiabatic regime.

The transport is accompanied  by a moderate heating of the cloud
(on the order of $40$ $\mu$K for both schemes) and no detectable
atom loss as soon as the transport duration is longer than two
periods of oscillation. For shorter times the acceleration is
sufficient to spill atoms out of the tweezers when $\eta=13$. The
increase of temperature is attributed to the transverse shaking of
the cloud that occurs during the transport. To circumvent this
limitation, we plan to use for future experiments an air bearing
translation stage instead of the standard linear rail guided
translation stage we are currently using. Note that the photon
scattering rate remains relatively small for both schemes (we
evaluate the photon scattering induced heating rate to be $3$
$\mu$K/s).

To infer the residual amplitude of  oscillation $\mathcal{A}$ (see fig.~\ref{fig.2}b and c), we measure the centre of mass oscillations after the transport by recording a set of typically $30$ images separated from one another by $10$ ms after the transport. The images are acquired using a standard absorption imaging technique on a CCD camera. Since the imaging process is destructive, the whole experimental sequence has to be redone for each picture. The position of the centre of mass of the cloud as a function of time is inferred from a $2$D Gaussian fit. We deduce the amplitude of oscillation by fitting the first period of this position data (see inset of fig.~\ref{fig.2}b) with a sine function.

For both schemes the variation of the amplitude as a function of
the transport duration is non monotonic. There are specific
discrete transport durations for which the measured amplitude of
oscillation is zero within our error bars (see second inset of
fig.~\ref{fig.2}b). This shows our ability to move a packet of
atoms in the non adiabatic regime (i.e. in a time on the order of
a few oscillation periods) with no excitation of the dipole mode
of oscillation {\it after} the transport. Note that the dipole
mode of oscillation is excited during the transport process since
the transport is non-adiabatic. We point out that after such an
optimal transport over a macroscopic distance, the number of atoms
and temperature of the remaining cloud are compatible with the
evaporative cooling to degeneracy in a crossed dipole trap
geometry. Indeed we have been able to achieve Bose Einstein
condensation with such clouds by crossing vertically a $200$
$\mu$m waist beam with our tweezers and ramping down both powers.

To interpret our data a simple one-dimensional analytical model is
sufficient and provides a good quantitative understanding of the
physics of the centre of mass motion of a packet of atoms
transported by a moving harmonic potential A similar formalism has
been developed to transport ions in segmented Paul trap
arrays~\cite{RLB06}. We consider an atomic packet initially at
rest in a harmonic trap of angular frequency $\omega_0$. The trap
position is given by the position of its centre $x_c(t)$. As
mentioned earlier, the movement of the centre of mass is decoupled
from the other degrees of freedom and hence can be treated as a
single particle in the harmonic trap. For a particle of mass $m$,
the imposed motion of the trap can be considered as an extra force
whose expression is $-m\ddot{x}_c(t)$ in the frame attached to the
trap.  According to Newton's law, the time dependent position
$x(t)$ of the centre of mass obeys the relation:
\begin{equation}
x(t) = x_c(t)+ \frac{1}{\omega_0} \int_{0}^{t}d t'
\sin[\omega_0(t'-t)]\ddot{x}_c(t') \ . \label{eq1}
\end{equation}
The amplitude $\mathcal{A}$ of the oscillatory motion after transport is readily inferred from eq.~\eqref{eq1}, and corresponds to the Fourier transform of the velocity profile of the trap's centre position:
\begin{equation}
\mathcal{A} = |{\cal F}[\dot{x}_c](\omega_0)| \ ,
\label{eq.amplitude}
\end{equation}
with ${\cal F}[f]=\int_{-\infty}^{+\infty} f(t){e}^{-{i}\omega t}\, dt$.

\begin{figure}
\onefigure{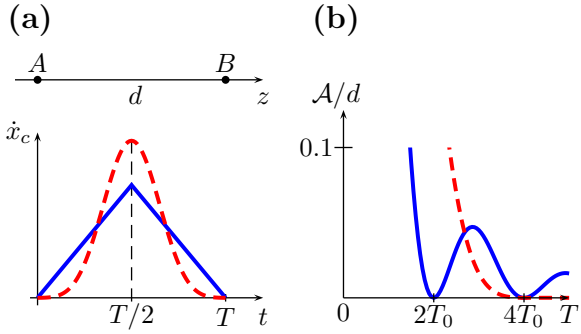} \caption{(Colour online). (a) Two different velocity profiles to go from a point $A$ to a point $B$ separated
by a distance $d$: the triangular profile (solid line) and the 4-term Blackman-Harris profile~\cite{HAR78} (dashed line).
(b) The residual amplitude $\mathcal{A}$ of oscillation of the centre of mass after transport for these velocity profiles (see text). An optimal transport ($\mathcal{A}=0$) can be performed in two period of oscillation in the triangle case, and in any time greater than $4T_0$ in the Blackman case.} \label{fig.1}
\end{figure}

In the case of a one-way transport over a distance $d=aT^2/4$ of duration $T$ with the simple velocity profile shown on fig.~\ref{fig.1}a (solid line), the final amplitude of oscillation is plotted on fig.~\ref{fig.1}b (solid line) and reads:
\begin{equation}
\mathcal{A}=d\ \mbox{sinc}^2(\omega_0 T/4) \ , \label{eq.AB}
\end{equation}
where the  $\mbox{sinc}(x)$ function is defined as $\sin(x)/x$. It exhibits a series of discrete optimal transport durations $T_n=2nT_0$, where $T_0=2\pi/\omega_0$ is the period of oscillation of the trap, and $n$ a non zero integer, for which the amplitude after the transport vanishes. They correspond to a transport without residual dipole mode excitation. We find, for this specific example, that it is possible to move optimally a packet of atoms on any distance $d$ on a time as short as twice the oscillation period. This is to be contrasted with the transport in the adiabatic limit ($\omega_0 T \gg 1$) for which the transport's duration is long compared to $T_0 = 2\pi/\omega_0$. We emphasize that these optimal strategies are robust against experimental uncertainties: indeed an error of $10\%$ on the transport duration $2T_0$ would lead to a residual amplitude of oscillation less than the one obtained when transporting ten times slower in a non optimal manner (in $21T_0$).

In the case of a ``round trip'', the amplitude of oscillation after a transport of duration $T$ reads:
\begin{equation}
\mathcal{A} = 2d\ \mbox{sinc}^2(\omega_0 T/8) |\sin(\omega_0 T/4)|
\ . \label{eq.ABA}
\end{equation}
As expected, we find optimal transport duration corresponding to a cloud stopped after the forward motion $A\rightarrow B$. Indeed the backward motion $B\rightarrow A$ is then optimal too, and we recover the $\mbox{sinc}^2$ factor obtained for the one-way transport. In addition we obtain another set of zeros (due to the $|\sin|$ factor) for which the cloud is not at rest after the forward move. In this case, the energy given to the cloud in the first half of the motion is removed during the second part due to the time symmetry of the trajectory around $T/2$.

The dashed line in fig.~\ref{fig.2}a is given by
eq.~\eqref{eq.ABA} rescaled by a factor of $0.6$ and is in good
agreement with our experimental data. The measured amplitude of
oscillation is smaller than the predicted one because in our
experiments the oscillation of the centre of mass is damped when
its amplitude is large (see first inset of fig.~\ref{fig.2}b).
This is due to the fact that the cloud explores potential region
far away from the minimum where non-linearities play an increasing
role. In this instance, particles have different periods of
oscillation depending on their energy, and the observed damping
results from the average taken over this spectrum of oscillation
frequencies involved in a transport experiment. It means that,
strictly speaking, it is impossible to transport in an optimal
manner a packet of atoms in the non-adiabatic regime as soon as
the potential exhibits non-linearities.

Two strategies can be used to avoid this effect. First, a longer transport time whilst remaining in the non-adiabatic regime minimizes this problem, because the cloud then remains close to the harmonic bottom of the trap. For scheme 1, we indeed observe that the damping is negligible for longer transport duration.

Alternatively, one can use a larger $\eta$ parameter. The involved
spectrum of oscillation frequency is then narrower, resulting in
partial damping suppression. This is exemplified by the data of
scheme 2 represented in fig.~\ref{fig.2}c for which $\eta = 50$
(to be compared to $\eta=13$ for scheme 1), where the dashed line
represents the theoretical prediction of eq.~\eqref{eq.ABA}
without any adjustment on the amplitude. For this sufficiently
large $\eta$, we recover the expected contrast of the amplitude
curve (see Figs.~\ref{fig.2}b and c). The simple theoretical
framework that we have developed is then in very good agreement
with our data.

The occurrence of optimal transport durations is a general feature
of the transport in the non-adiabatic regime with a harmonic trap.
They can be adjusted at will by choosing a proper velocity profile
for the displacement of the trap. The duration of an optimal
transport can in principle be reduced to very short time in
comparison to the period of oscillation. However for practical
reasons, including the finite depth of the trapping potential,
there is always a limit on the acceleration one can use and thus
on the minimum transportation time.

The Fourier transform formulation (eq.~\eqref{eq.amplitude}) of
the transport allows for many enlightening analogies. For
instance, the modulus square of the amplitude $\mathcal{A}^2$ is
mathematically identical to the intensity profile for the far
field Fraunhofer diffraction pattern of an object with a
transmittance having the same shape as the velocity profile for
the transport. An optimal transport condition is equivalent to a
dark fringe in the corresponding diffraction pattern. The ``round
trip'' $A\rightarrow B \rightarrow A$ (see fig.~\ref{fig.2}a)
considered in our experiment is made of two triangular velocity
profiles corresponding to a one way transport $A \rightarrow B$
and another in the opposite direction $B \rightarrow A$. In
optics, we know that the repetition of a pattern in the
transmittance yields interferences. We can thus re-interpret the
formula~\eqref{eq.ABA} where the factor term
$\mbox{sinc}^2(\omega_0 T/8)$ plays the role of a diffraction
pattern for a one way transport, and the factor term
$|\sin(\omega_0 T/4)|$  accounts for ``interferences'' between the
two one-way velocity profiles. The optimization of the conditions
under which a non adiabatic transport should be carried out with a
``harmonic'' optical tweezers are then equivalent to apodization
problems in optics, which is a thoroughly studied
problem~\cite{JAR64}. We further emphasize that the finite width
of the spectrum of periods of oscillation due to non linearities
involved in a non-adiabatic transport is reminiscent of the finite
temporal coherence of the illuminating source in an optical
diffraction experiment. In the case of transport the width of this
spectrum depends on the duration of the transport. In optics the
spectrum width is an intrinsic property of the illuminating source
and therefore affects globally the whole diffraction pattern.

Another interesting analogy based on the Fourier transform formulation lies in the minimization of side lobes of the spectrum when choosing a window to perform spectral analysis, or when choosing the time shape of a Raman pulse~\cite{HAR78,KAC92}. For instance, the use of a 4-term Blackman-Harris shape for the velocity profile\footnote{$\dot{x}_c(t)=0.35875-0.48829\cos(2\pi t/T)+0.14128\cos(4\pi t/T)-0.01168\cos(6\pi t/T)$} should ensure a robust optimal transport as soon as its duration is bigger than $4T_0$ (see fig.~\ref{fig.1}, dashed lines), yielding a very robust optimal transport.

In conclusion, we have demonstrated the implementation of an optimal transport with optical tweezers in the non-adiabatic regime along with a simple theoretical formalism. The results presented in this letter are of interest not only for cold atoms experiments to increase their repeating rate, but also to any experiment where transport using harmonic traps is achievable like for instance trapped ions experiments.

\acknowledgments

We thank J. Dalibard and T. Lahaye for careful reading of the manuscript. Support for this research came from the D\'el\'egation G\'en\'erale pour l'Armement (DGA, contract number 05-251487), the Institut Francilien de Recherche sur les Atomes Froids (IFRAF) and the Plan-Pluri Formation (PPF) devoted to the manipulation of cold atoms by powerful lasers. G.~R. acknowledges support from the DGA.

\end{document}